# Game Design with Pocket Code: Providing a Constructionist Environment for Girls in the School Context

**Anja Petri,** *anja.petri@ist.tugraz.at*
Institute of Software Technology, Graz University of Technology, Austria

**Christian Schindler,** *christian.schindler@ist.tugraz.at*
Institute of Software Technology, Graz University of Technology, Austria

**Wolfgang Slany,** *wolfgang.slany@tugraz.at*
Institute of Software Technology, Graz University of Technology, Austria

**Bernadette Spieler,** *bernadette.spieler@ist.tugraz.at*
Institute of Software Technology, Graz University of Technology, Austria

## Abstract

The widespread use of mobile phones is changing how learning takes place in many disciplines and contexts. As a scenario in a constructionist learning environment, students are given powerful tools to create games using their own ideas. In the "No One Left Behind" (NOLB) project we will study through experimental cycles whether the use of mobile game design has an impact on learning, understanding, and retention of knowledge for students at risk of social exclusion. We will use the mobile learning app Pocket Code with partner schools in three countries: Austria, Spain, and the UK. This paper focuses on the Austrian pilot, which is exploring gender inclusion in game creation within an educational environment. We first study differences in game creation between girls and boys. This study that started in September 2015, will help teachers to integrate Pocket Code effectively into their courses. For future studies an enhanced school version of Pocket Code will be designed using the results and insights gathered at schools with pupils and teachers.

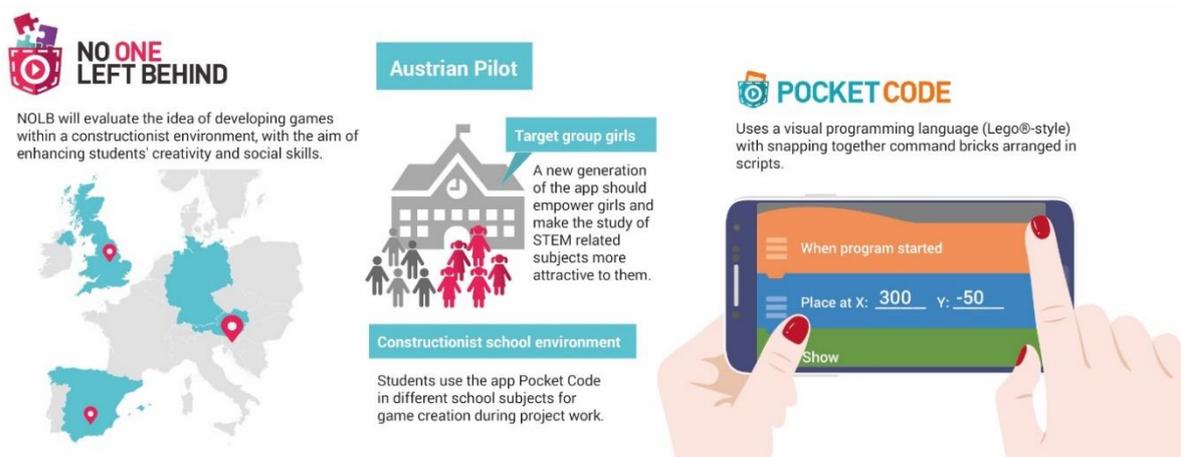

Figure 1: The NOLB project, the Austrians pilot and Pocket Code.

# Introduction

Constructionism motivates learning through creating one's own games (Parmaxi and Zaphiris, 2014) but requires appropriate hardware infrastructure, which is often outdated and/or insufficiently available in schools. Modern smartphones are increasingly owned by students all over the world and thus could help solve the hardware problem in and outside of schools. Playing mobile games is a popular leisure activity for young people (Schippers and Mak, 2015), but creating smartphone apps has until now been difficult. To ensure a constructionist learning environment, Parmaxi and Zaphiris recommended the use of appropriate mobile tools that can address the cognitive and social aspects of learning. Therefore we will support students creating their own mobile games in a visual, Lego®-style way to improve their learning experience at schools. With the mobile learning app Pocket Code students can easily program games directly on their smartphones, without requiring any additional hardware. Within the EU project "No One Left Behind" (NOLB), we will develop a constructionist approach that integrates Pocket Code into school curricula and that motivates students to become independent thinkers.

In previous work, the authors described reasons why girls and women are underrepresented in Science, Technology, Engineering, and Mathematics (STEM) related subjects (Beltrán *et al.*, 2015). In addition, Weibert *et al.* (2012) defined personal attachment as a determining factor to motivate girls to attend computer science subjects. Therefore, the challenge of a constructionist learning environment is to create situations in which girls feel comfortable expressing their own interests. Within NOLB, groups of girls and boys will solve curriculum related problems while using the Pocket Code app. When working on these projects, students can gather information from a variety of sources, analyse this information, engage with it intensely, and derive knowledge from the process of constructing a game (Chandrasekaran, 2012). This connection with real world problems enhances their learning and makes it more valuable. Thus game design methods utilise game and project based learning and foster collaboration for students at risk of social exclusion.

This paper starts by describing gender differences in game design, followed by background information about constructivism and constructionism. The next section provides details about the research methods used within the NOLB project, the Pocket Code app, the Austrian pilot, and goals of the first feasibility study. The next sections illustrates the integration of Pocket Code in the school contexts and describes workshops with students and teachers in Austria. The intent is to connect students' interests with coding while at the same time fostering a constructionism approach through collaboration and engagement to make learning enjoyable. Clearly it is important to support students and teachers with material and to provide resources such as media assets and frameworks. Finally, we describe possible future directions for research.

# Related work

In this section we describe studies about gender aspects in game play and design. We also present settings for a constructivist and constructionist school environment.

## Supporting gender unbiased game design

In 1996, Greenfield and Cocking discovered that there are significant gender differences between girls and boys regarding performance, interests, and experience in game play. Kafai (2008) proposes to emphasize different content, mechanics, and characters in order to make games more appealing to girls. Craig *et al.* (2013) observed that the participation of girls in computer classes is not the same as those of boys: girls tend to spend more time on visual customization while boys spent more time on solving logical puzzles, and the authors thus point out that it is essential to consider gender differences in logical and computational skills. According to this study it may thus be more effective to get girls interested in technology by asking them to design games rather than to focus on the learning of specific programming skills.

## Constructivism and constructionism

Piaget's constructivism provides a framework for optimizing the learning progress at different levels of children's development (Kafai and Resnick, 1996). Younger children create their own subjective reality, depending on their own experiences, which is suited to their current needs and possibilities. Children enhance their capability of abstract thinking and start to philosophize about probabilities, associations, and analogies by the age of 11.

Papert's constructionism is the practical realization of the constructivism theory. Papert (1980) noted that individual learning occurs more effectively when students understand the world around them by creating something that is meaningful to themselves. These can be artefacts such as a sand castle or a computer program (Parmaxi and Zaphiris, 2014). Programming tools such as Scratch[1] make programming accessible to a large number of people and teach new skills and abilities, such as engineering, design, and coding (Blikstein and Krannich, 2013).

# Research setting

The NOLB project plans to validate its approach by conducting three pilot studies from January 2015 to June 2017 in Austria, the UK, and Spain. In each of these pilots participate approximately 200 students (including the control groups) between the ages of 9 and 18 years to study different social exclusion problems. The framework for this study refers to the theory of constructionism, which emphasizes design and sharing of artefacts (Parmaxi and Zaphiris, 2014). This section describes the NOLB project, the Pocket Code app, the research setting of the Austrians pilot, and the feasibility study.

## The "No One Left Behind" project

The "No One Left Behind" (NOLB) project[2] aims at unlocking game creation by students who are socially excluded. Students will use Pocket Code to develop games on mobile devices, with the goal of enhancing their abilities across all academic subjects as well as their computational proficiency, creativity, and social skills. The pilots address three inclusion challenges: gender exclusion, disabilities, and immigration. We plan to use a constructivism approach by learning through Pocket Code. This app will support students in constructing their own games in the context of their regular courses, by embedding academic elements in their games and reflecting their understanding of what they need to accomplish. Moreover, students can socialize with their peers during the game making process. The results will include designed sharable artefacts that reflect the students' different styles of thinking and learning (Ramnarine-Rieks, 2012).

## Pocket Code

Pocket Code is an application for mobile devices and tablets. It is currently available for Android (a version for iOS and Windows Phone is in development). This app provides a visual programming language which allows students to create their own games, animations, interactive music videos, and many types of other apps, directly on their smartphones or tablets (Catrobat, 2015). Similar to Scratch, programs in Pocket Code are created by visually composing Lego®-style bricks arranged in "scripts" which can run in parallel, thereby allowing concurrent execution.

## Setup in Austria: Empowering girls' computational skills

The Austrian pilots are conducted together with three high schools situated in and around Graz: "Akademisches Gymnasium Graz", "Graz International Bilingual School" (GIBS), and "BORG Birkfeld". 190 students from 11 to 16 years, of which 122 are girls, take part in the study. The intent is to find out how to develop Pocket Code in ways that specifically empower girls by engaging with them and thus making the study of STEM related subjects more attractive to them. Additionally, the differences in the game creation process will be analysed. Aspects of self-identity and stereotyped gender categories will be taken into consideration.

---

[1] https://scratch.mit.edu/
[2] http://no1leftbehind.eu

## Feasibility Study: Integrating Pocket Code in school lessons

Figure 2 outlines the different stages of the feasibility study. Pocket Code is used in several academic subjects, thus the purpose of this study is twofold. First, the feasibility pilot allows students and teachers to familiarize themselves with Pocket Code through applying game design elements in selected curricula areas. By this we will identify students' and teachers' needs regarding the app. Second, the study gives students an initial positive experience with the app. Pocket Code will allow them to engage with their subjects in a playful way.

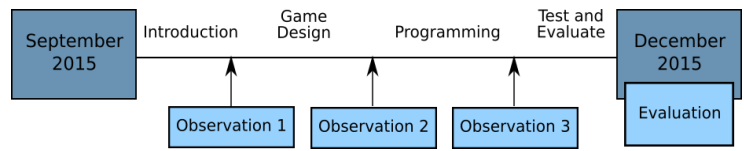

*Figure 2. Schedule of the feasibility study.*

The results of the study will shape the new version of Pocket Code through the feedback from students and teachers. At the end of the study the following evaluations are planned:

- Analysis of data collected (pre- and post-surveys, observations during the study).
- Defining the range of capabilities measured through gaming analytics and the progression of the learners.
- Learning about the barriers of using Pocket Code in schools.
- Measuring the learning objectives against the learning outcomes.
- Documenting findings to improve future studies and developments.
- Reviewing school capabilities and issues.
- Exploring the role of the teachers (teachers as facilitators, coaches).

Additionally, the following aspects are specific to the Austrian study:
- Differences in game creation and design between girls and boys.
- Suggested changes that are needed to make Pocket Code more attractive to girls.

# Preparation phase and feasibility study in Austria

This section describes the preparation phase in Austrian pilot schools and provides a summary of the workshops with students as well as an overview of the teacher training sessions and prepared material. We further describe how findings from the preparation phase have influenced the ensuing feasibility study and the students' lessons with Pocket Code.

## Pocket Code workshops with students

At the end of the school semester in July 2015, we organised workshops in two of our partner schools, with a focus on introducing Pocket Code to students. The theme for the workshop was "150 years of Alice in Wonderland" in reference to the 150th anniversary of Lewis Carroll's book celebrated in 2015. This topic seemed interesting to the target audience: students 14-17 of both genders. For the workshop itself, various materials were prepared, including media assets such as graphics, sounds, and tutorial cards for important functions (e.g., set size of an object). One NOLB partner, the National Videogame Arcade (NVA) in the UK, has developed a taxonomy called "The Shape of a Game" ceremony. It consists of a title screen, an instruction screen, a game screen, and an end screen. This framework was preinstalled on the mobile phones the students used for the workshop. With this framework the students' game design processes were scaffold, allowing them to focus on the game development itself. At GIBS 19 and at Borg Birkfeld 35 students attended the four hour workshop.

The introduction session was held in front of the whole class to show what could be achieved within Pocket Code, with the team presenting example games and the user interface. Afterwards the students formed groups of two or three and pitched game ideas. These ideas were shared with the class and the students got direct feedback. Next, the students created storyboards which helped them to get a clearer image of the gameplay and characters. Without giving them further guidance in programming, they started to code their games. The team took the role of coaches, and students could consult them when they needed some input. This learning by doing approach supports the constructionist theory, and most importantly students had the opportunity to add their creativity to this session and explore the various functionalities of Pocket Code on their own. Figure 3 illustrates different impressions of the workshop. At the end they uploaded their games to our web share page[3] and presented them in front of their peers.

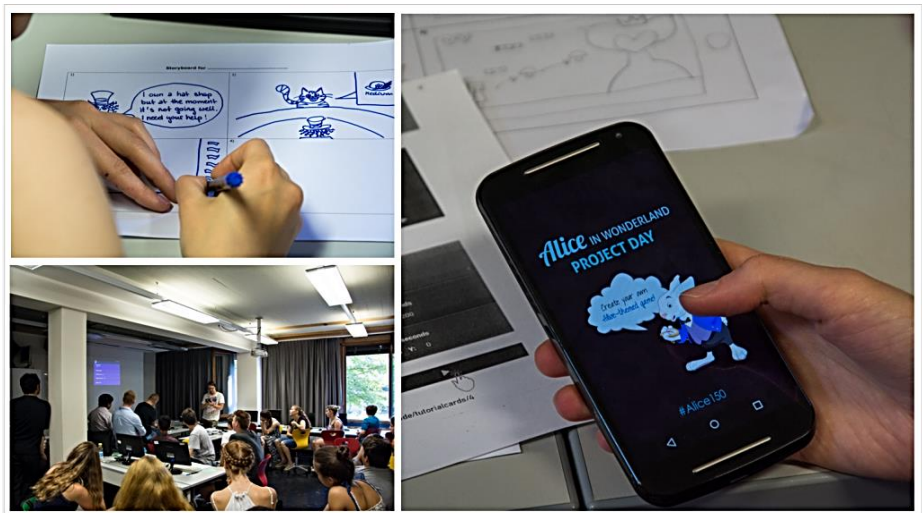

*Figure 3. Pictures of the workshop: storyboard, instruction session and framework "Shape of a Game".*

In both workshops many different games were created, e.g., mazes, skill games, jump and run games, quizzes, adventure games, or racing games. We conducted a survey at the end of the workshops. The results showed that 98% of the students described the workshop as very good or good, 90% were satisfied with their results, and 89% liked to work in teams. While observing the students use of Pocket Code, challenges could be identified, e.g., working in groups or difficult parts while programming. The students had especially problems in creating the following functions: to detect a collision, to move the background, and to use variables. Furthermore, the games created have been analysed in order to define what kind of game design elements were used most often and to generate therefore tutorials cards, tutorial videos and frameworks with Pocket Code. Additionally, the workshop provided a good opportunity to gain more experience in applying game oriented methods within the curricula.

## Pocket Code teacher training sessions

Before finalising the planning phase, the teachers needed to be involved as well. The success of the project depends on their participation and cooperation. Influencing factors include the subjects in which teachers will use Pocket Code and the amount of units available. Preliminary work included several meetings with teachers in spring 2015 and conducting an online questionnaire to establish the teachers' digital skills and abilities. Initial results were used to develop the classroom resources that teachers need to implement, based on their differentiated skill levels. In September 2015, the first teacher trainings were held to show the functionalities of Pocket Code. In the two hour courses, teachers were given a short introduction to Pocket Code, after which they had a hands-on session in form of a step-by-step approach with the workshop facilitator. The sessions were followed by short discussion rounds where teachers considered how Pocket Code could be used best to support their lessons. Afterwards the teachers created some excellent ideas together with their students. In the following weeks they sent these ideas either in form of storyboards or description to the NOLB team and got feedback from us. After both the teacher trainings and the student workshops, the following materials were prepared to kick off the feasibility study:

- Online course[4] including explanation of the bricks, tutorial cards for students, and helpful game metrics (shape of a game, game design process)
- Short video tutorials with gaming concepts[5]
- Pocket Code frameworks for Physics, Arts, English, Computer Science and Music[6]
- Media assets like graphics and sounds online through the Pocket Code Media Library

---

[3] https://pocketcode.org/

The effectiveness and usefulness of this material will be tested during this feasibility study.

## Kick-off of the feasibility study

With the help of the NOLB team the teachers will use different approaches depending on the age of their students, their previous programming experience and their subject field. Table 1 points out the various approaches which all have a constructionist base in common e.g., students will be actively involved in the process of constructing a game, they will take charge of their own learning and find solutions by using the tutorial cards, asking their peers or just trying out alternatives and learn from their mistakes. Therefore, students will learn to be patience, feel ownership of their achievements and will truly understand many of the basic programming steps. To archive this, students will work in at least three units; in the Austrian pilots, girls will be grouped together. For the purpose of designing and creating artworks for backgrounds and objects we will provide students enough time during additional Arts lessons (important for girls see previous section).

| Pilot school 1: "Akademisches Gymnasium" | |
|---|---|
| Arts (2nd grade) | Idea: 17 different languages are spoken in this class, thus the idea is to create a vocabulary game. <br> Game play: A list with different vocabulary words is shown on the screen. In the game, balloons with pictures glide. The player should try to catch the balloons that show vocabulary words from the list using the inclination sensors of the phone. A score displays changes with right/wrong catches, and when all right balloons are caught the game is finished. The students should add functionalities, own graphics, and other languages/vocabulary to their games. |
| Computer Science (5th grade) | Idea: In these units the students will not receive a given project theme. The aim is that they will be able to work creatively and to explore the basics of a programming language on their own. The sequence of this lesson will be similar to the workshop: They will create a storyboard, present their ideas, and start programming. |
| Pilot school 2: "GIBS" | |
| Physics (3rd grade) | Idea: This game should deal with the density of different objects and liquids. The lessons should follow up with performing these experiments in real life. <br> Game play: The game shows the behaviour of different objects (e.g., a bolt, a cork) falling in different liquids (e.g., water, oil). The students should add further objects (e.g., paper, coins) and liquids (e.g., honey). |
| English & Computer Science (3 classes 5th grade) | Idea: The idea is to program a quiz game with the students. In the English class the students had to read five different books and can choose which one of them they will use for the quiz. <br> Game play: In the Computer Science class students will program these quizzes. At five different stations questions will be asked concerning the book, followed by a small mini game. If all questions are answered correctly, the game is finished. The students should create the stations with questions and mini games. |
| Arts (5th and 6th grade) | Idea: This teacher will use a different approach. He will use the brick cards the NOLB team created, print them out, and use them as a physical flashcard system to put together scripts. Thus students draw first the projected outcome on a picture. When they start programming their games in Pocket Code, they can see if it follows the same configuration that they predicted. They may choose the topic of their games freely. |
| Pilot school 3: "Borg Birkfeld" | |

| Computer Science (5th grade) | Idea: This class will do also a quiz game in Pocket Code and answer computer related questions.<br>Game Play: The player listens to a question, e.g., "Catch all object-oriented programming languages" and then catch the correct image from several that are falling. A score display changes with right/wrong catches. When all questions are answered correctly the game is finished. The students should add additional questions to the game. |
|---|---|
| Music (5th grade) | Idea: Same concept as in Computer Science.<br>Game Play: The player listen to music, e.g., "The Magic Flute" and then the player should catch the instruments that were part of the music. The students should add additional music and instruments. |

*Table 1. Ideas for the students' lessons in the different pilot schools.*

Students of the 2nd grade will get a framework in which specific parts in the code are missing (indicated by a note brick): e.g., collision detection or using inclination sensors. Within small groups the team develops this missing functionality together with them to guarantee student-centered classroom, where students learning and discovery is in their own hands. Students up to 3rd grade will get a short introduction unit including a hands-on sessions guided by the team and tutorial cards. In these sessions one student at a time comes to the front of the class and tries to add one small but meaningful new feature to the game being developed by the class.

Pre and post questionnaires, observations, documentations, video, and picture material will be collected during all units. The outcomes of the feasibility study will be incorporated in the next planned cycles. These findings will be integrated in a new generation of Pocket Code, which will be designed and implemented especially for school contexts. For the first cycle in summer semester 2016 we will continue with the same classes that are already experienced in Pocket Code. The new version will be tested and evaluated during the second cycle, which will start in September 2016. In order to monitor these cycles, NOLB will select an experimental design (experimental vs. control group).

# Future work

Future work will include the testing of more playable approaches to aid girls in certain topics. One way will be through Game Jams events. Goddard *et al.* (2014) see Game Jams as a way to generate and inspire game ideas and finding new ways of thinking. Thus NOLB will examine the opportunity of utilizing Game Jams as a design research tool in school classes. One event (at the time of this writing still upcoming) which will be also interesting for schools is the Alice Game Jam[7] in December 2015.

A series of data collection tools (teachers' questionnaire, student gaming questionnaire, and baseline collection tool) have been designed for understanding students' academic, social and inclusion needs, requirements and capabilities. They will help to identify students' and teachers' needs to guide the development of an inclusive generation of Pocket Code, influence the design of the full experimental pilots and provide benchmarks against which we can evaluate the system's potential in generating inclusive game creation experiences in formal learning situations.

# Discussion and conclusion

Many of the changes in teaching and learning that resulted from the study of empowerment of girls improved the situation for all students, not just for girls (Kafai, 2008). Within NOLB it is assumed that learners who become game designers and creators will significantly contribute to closing the divide and participation gap in digital culture. Preliminary findings show that students were able to successfully design functional games in a very short time frame using Pocket Code. A more detailed analysis will be available after the feasibility study, the findings of which will help to understand the game design behaviour of girls and to identify obstacles in usage of Pocket Code.

---

# Acknowledgements

This work has been partially funded by the EC H2020 Innovation Action No One Left Behind, Grant Agreement No. 645215.

# References

Beltrán, M. *et al.* (2015) *Inclusive gaming creation by design in formal learning environments: "girly-girls" user group in No One Left Behind*, In Design, User Experience, and Usability: Users and Interactions. Los Angeles. July, pp. 153 – 161.

Blikstein, P. and Krannich, D. (2013) *The Makers' Movement and FabLabs in Education: Experiences, Technologies, and Research.* Proceedings of the 12th International Conference on Interaction Design and Children. New York. June, pp. 613 – 616.

Craig, A.; Coldwell-Neilson, J. and Beekhuyzen, J. (2013) *Are IT interventions for Girls a Special Case?* In Proceeding of the 44th ACM technical symposium on Computer science education. Denver, March. pp. 451 – 456.

Chandrasekaran, S. (2012) *Learning through Projects in Engineering Education.* SEFI 40th annual conference. Greece. 2012

Goddard, W.; Byrne, R. and Mueller, F. (2014) Playful Game Jams: Guidelines for Designed Outcomes. Proceedings of the 2014 Conference on Interactive Entertainment. Newcastle. December, pp. 1 – 10.

Greenfield, P. M. and Cocking, R. R (1996). *Interacting with video.* Norwood, NJ: Ablex (expanded version of Greenfield & Cocking, 1994).

Kafai, J. (2008) *Considering Gender in Digital Games: Implications for Serious Game Designs in the Learning Sciences.* In Proceedings of the 8th international conference on International conference for the learning sciences. Los Angeles, pp. 422 – 429.

Kafai, Y. and Resnick, M. (1996): *Constructionism in Practice: Designing, Thinking, and Learning in a Digital World.* Mahwah, New Jersey, Lawrence Erlbaum.

Papert, S. (1980) *Mindstorms: Children, computers, and powerful ideas.* Basic Books, Inc.

Parmaxi, A. and Zaphiris, P. (2014) *Affordances of Social Technologies as Social Microworlds.* Extended Abstracts on Human Factors in Computing Systems. Toronto, April. pp. 2113 – 2118.

Ramnarine-Rieks, A. (2012) *Learning through Game Design: An Investigation on the Effects in Library Instruction Sessions.* Proceedings of the 2012 iConference. Toronto. Feb., pp. 606 – 607.

Schippers,S. and Mak. M (2015) *Creating Outstanding Experiences for Digital Natives. A survey of Digital Natives reveals a group of impatient users with fragmented attention spans who demand fast and intuitive products and services.* Article No: 1278. Retrieved from: http://uxmag.com/articles/creating-outstanding-experiences-for-digital-natives.

Sanchez, J. and Olivares, R. (2011) *Problem solving and collaboration using mobile serious games.* In Computers & Education 57(3). Santiago. November, pp. 1943 – 1952.

Catrobat (2015) *Catrobat.* Last visited: October, 6th 2015. Retrieved from: http://www.catrobat.org.

Weibert, A., Rekowski, T. and Festl, L. (2012) *Accessing IT: a curricular approach for girls.* In Proceedings of the 7th Nordic Conference on Human-Computer Interaction: Making Sense Through Design. Copenhagen, October. pp. 785 – 786.

BibTex entry:

```
@InProceedings{PETRI2016APP,
author = {Petri, P. and Slany, W. and Schindler,  C., and Spieler, B. }
title = {Game Design with Pocket Code: Providing a Constructionist
Environment for Girls in the School Context}
booktitle = {Proceedings of the 4th Conference on
Constructionism},
isbn = {978-616-92726-0-1},
location = {Bangkok, Thailand},
month = {06-07 February, 2016},
year = {2016},
pages = {109-116}}
```